\documentclass[prd,aps,showpacs,,nofootinbib,superscriptaddress,preprintnumbers]{revtex4}
\usepackage[dvips]{graphicx}

\begin{document}
\preprint{\hbox{RUB-TPII-04/08}}

\title{Duality between different mechanisms of QCD factorization
       in $\gamma^*\gamma$ collisions}

\author{I.~V.~Anikin}
\email{anikin@theor.jinr.ru}
\affiliation{Bogoliubov Laboratory of Theoretical Physics, JINR,
             141980 Dubna, Russia}
\author{I.~O.~Cherednikov}
\email{igor.cherednikov@jinr.ru}
\affiliation{Bogoliubov Laboratory of Theoretical Physics, JINR,
             141980 Dubna, Russia}
\affiliation{Institut f\"{u}r Theoretische Physik II,
             Ruhr-Universit\"{a}t Bochum,
             D-44780 Bochum, Germany}
\affiliation{INFN Gruppo collegato di Cosenza, I-87036 Rende, Italy}
\author{N.~G.~Stefanis}
\email{stefanis@tp2.ruhr-uni-bochum.de}
\affiliation{Institut f\"{u}r Theoretische Physik II,
             Ruhr-Universit\"{a}t Bochum,
             D-44780 Bochum, Germany}
\author{O.~V.~Teryaev}
\email{teryaev@theor.jinr.ru}
\affiliation{Bogoliubov Laboratory of Theoretical Physics, JINR,
             141980 Dubna, Russia}
\date{\today}

\begin{abstract}
We study the phenomenon of duality in hard exclusive reactions
to which QCD factorization applies.
Considering ``two-photon''-like processes in the scalar $\varphi^3_E$
model and also two-hadron (pion) production from the collisions
of a real (transversely polarized) and a highly virtual, longitudinally
polarized, photon in QCD, we identify two regimes of factorization
each of them associated with a distinct nonperturbative mechanism.
One mechanism involves twist-$3$ Generalized Distribution Amplitudes,
whereas the other one employs leading-twist Transition Distribution
Amplitudes.
In the case of the scalar $\varphi^3_E$ model, we find duality in
that kinematical region where the two mechanisms overlap.
In the QCD case, the appearance of duality is sensitive to the
particular nonperturbative model applied and can, therefore, be used
as an additional adjudicator.

\end{abstract}

\maketitle

\section{Introduction}
\label{sec:intro}

Lacking a complete theoretical understanding of {color
confinement, the only method of applying QCD is based on the
factorization of the dynamics and the isolation of a short-distance
part which is amenable to perturbative techniques of quantum field
theory
(see, \cite{Efremov-Radyushkin,Bro-Lep,Col-Sop-Ste89} and for
a review, for instance, \cite{Ste99} and references cited therein).
On the other hand, the long-distance part has to be parametrized
in terms of matrix elements of quark and gluon operators between
hadronic states (or the vacuum).
These matrix elements have nonperturbative origin and have to be
either extracted from experiment or be determined by lattice
simulations.
In many phenomenological applications they are usually modeled
by applying various nonperturbative methods.

The QCD description of more involved hadronic processes requires
the introduction of new hard parton subprocesses and nonperturbative
functions.
Important particular examples are exclusive hadronic processes which
involve hadron distribution amplitudes (DAs), generalized
distribution amplitudes (GDAs), and generalized parton
distributions (GPDs) \cite{Diehl:2003,Bel-Rad,Ji,NonforRad,GPV}.
Within this context, collisions of a real and a highly-virtual
photon provide a useful tool for studying a variety of fundamental
aspects of QCD.
A special case of this large class of processes is the exclusive
two-hadron (pion) production in that region, where one initial
photon is far off-shell (with its virtuality being denoted by $Q^2$)
and longitudinally polarized, with the other photon being transversely
polarized, while the overall energy (or, equivalently, the invariant
mass of the two hadrons), $s$, is small.
Such a process factorizes into a perturbatively calculable
short-distance dominated subprocess, that describes the scattering
$\gamma^* \gamma \to q \bar q$ or $\gamma^* \gamma \to g g$,
and nonperturbative matrix elements measuring the transitions
$q \bar q \to A B$ and $g g \to AB $.

Another extensively studied---both theoretically and
experimentally---two-photon process is the pion-photon transition
form factor.
This exclusive process has been measured by the CELLO \cite{CELLO91}
and the CLEO \cite{CLEO98} Collaborations for an almost on-shell
(i.e., real) photon, whereas the other one has a large virtuality of
up to 9~GeV${^2}$ and is transversely polarized.
These data have been analyzed over the years by many authors using
various theoretical approaches---see, for instance,
\cite{KR96,RR96,SSK99,SSK00,SY99,DKV01,BMS02,BMS03,BMS05lat,Ste08}.
The major outcome of the most recent study \cite{BMS05lat}
is that the CLEO data exclude the Chernyak-Zhitnitsky pion distribution
amplitude \cite{CZ84} at the $4\sigma$ level, while, perhaps somewhat
surprisingly, also the asymptotic distribution amplitude seems to be
disfavored (being off at the $3\sigma$ level).
These findings are supported by the latest high-precision lattice
simulations by two independent collaborations \cite{Lat06,UKQCD-RBC07}.
It is remarkable that the only pion distribution amplitude found to be
within the $1\sigma$ error ellipse of the CLEO data, while complying
with the new lattice results just mentioned, is the pion model
derived from nonlocal QCD sum rules \cite{BMS01}.

These results lend credibility to the usefulness of two-photon
processes as a means of accessing and understanding the parton
structure of hadrons.
On that basis, one may try to assess more complicated two-photon
processes with two pions in the final state.
Indeed, recently, nonperturbative quantities of a new kind were
introduced---transition distribution amplitudes
(TDAs) \cite{Frank-Pol,Pire-Szym,LPS06}---which are closely related to
the GPDs and describe the transition $q \gamma \to q A$.
In contrast to the GDAs, mentioned above, such nonperturbative
objects, like the TDAs, appear in the factorization procedure
when the Mandelstam variable $s$ is of the same order of magnitude
as the large photon virtuality $Q^2$, while $t$ is rather small.
Actually, there exists a reaction where both amplitude types,
GDAs and TDAs, can overlap.
Precisely this can happen in the fusion of a real transversely
polarized photon with a highly-virtual longitudinally polarized
photon that gives rise to a final state comprising a pair of pions.
This reaction can potentially follow either path, i.e., it can
proceed via the twist-$3$ GDAs, or through the leading-twist TDAs, as
illustrated in Fig.\ \ref{gen-fact-mechs}.
The first mentioned possibility is linked to the case when one of the
involved photons is longitudinally polarized, whereas the other one is
transversely polarized.
The second possibility corresponds to the same situation, but appears
in contrast to that in leading-twist order, as we will show in Sec.\
\ref{sec:fact2}.
Thus, a comparative analysis of both possible mechanisms for this
reaction seems extremely interesting both theoretically and
phenomenologically.

A simultaneous analysis of these two mechanisms in the production of
a vector-meson pair was carried out in \cite{PSSW}.
The authors found that these mechanisms can be selected by means of
the different polarizations of the initial-state photon.
In the case of (pseudo)scalar particles, like the pions, this effect
is absent and, therefore, this opens up a window of opportunity to
access the overlap region of both mechanisms and their duality
as opposed to their additivity.
By duality we mean here that adding the contributions of the two
mechanisms would lead to a double-counting.
Hence, these mechanisms are in antagonism to each other and
should be considered as two alternative ways to describe the same
physics.

In this paper, we will verify the possibility for duality
between the two different mechanisms of factorization, associated
either with GDAs, or with TDAs, in the regime where \textit{both}
Mandelstam variables $s$ and $t$ are rather small compared to the
large photon virtuality $Q^2$.
This will be done below by considering first (Sec.\ \ref{sec:fact1})
the Euclidean $\varphi^3_E$-analogue of QCD, followed then by an
investigation of the exclusive pion-pair production in
$\gamma\gamma^*_L$ collisions in Sec.\ \ref{sec:fact2}.
Our findings are summarized and further discussed in Sec.\
\ref{sec:concl}, where we also provide our conclusions.
\begin{figure}[t]
\centerline{\includegraphics[angle=90,width=0.6\textwidth]{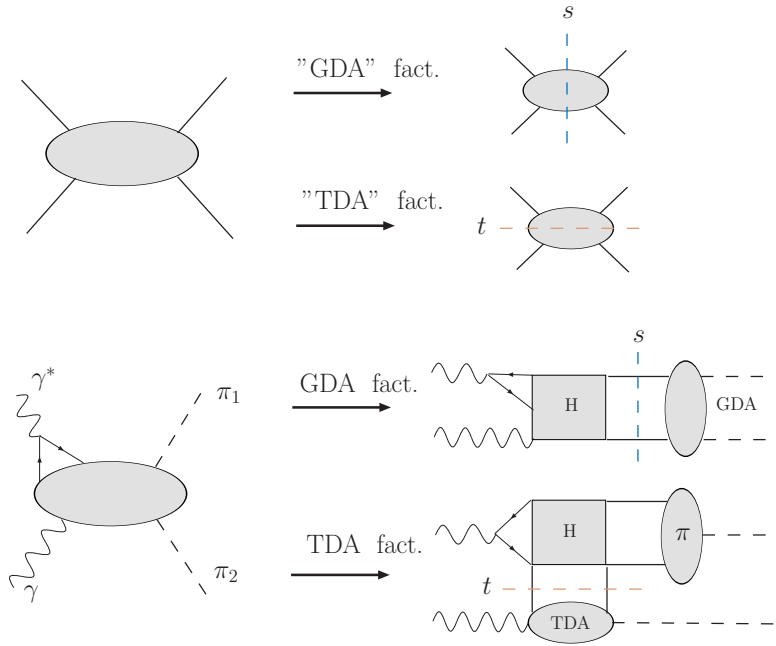}}
  \caption{(Top) Dual factorization mechanisms in describing the
  fusion of a real photon with a highly-virtual and longitudinally
  polarized photon in the Euclidean $\varphi^3_E$ model.
  (Bottom) The analogous situation in QCD.
   \label{gen-fact-mechs}}
\end{figure}

\section{Regimes of Factorization within the $\varphi^3_E$-model}
\label{sec:fact1}

We start our analysis with the factorization of a two-photon process
in which the two photons undergo a fusion with the subsequent
creation of two pions.
For the sake of simplicity, we work first in the scalar Euclidean
$\varphi^3_E$ model, regarding it as a sort of a toy model in order
to prepare the ground for the real QCD application to follow in the
subsequent section.

The most convenient way to analyze this four-particle amplitude in the
$\varphi^3_E$ model is to work within the $\alpha$-representation,
following the lines of thought given in \cite{NonforRad} in the
discussion of Deeply Virtual Compton Scattering.
The reason is that within this representation (i) the calculations
can be performed in a systematic way, not only at one loop but also at
higher-loop levels, (ii) the factorization of the process can be
studied in detail, and (iii) one can study the spectral properties of
the nonperturbative input (GPDs, GDAs) because these properties are
insensitive to the numerators of the quark and gluon propagators.
In addition, complications owing to the spin structure do not affect
them.
It is worth recalling in this context that the $\alpha$-representation of
Feynman diagrams is not only a schematic means for identifying relevant
integration regions in a factorization procedure, but can, in fact,
serve to provide rigorous proofs of factorization theorems
\cite{Efremov-Radyushkin}.

The contribution of the leading ``box'' diagram, see Fig. \ref{Fig01},
can be written as
\begin{eqnarray}
\label{Amp1}
    {\cal A}(s,t,m^2)
    =-\frac{g^4}{16\pi^2}
    \int\limits_{0}^{\infty} \frac{\prod\limits_{i=1}^4 d\alpha_i}{D^2}
    \text{exp} \biggl[ - \frac{1}{D} \left( Q^2 {\alpha_1\alpha_2}
    + s \alpha_2\alpha_4 +
    t {\alpha_1\alpha_3} + m^2 D^2 \right)\biggr],
\end{eqnarray}
where $m^2$ serves as a infrared (IR) regulator, $s>0$, $t>0$ are the
Mandelstam variables in the Euclidean region, and
$D=\alpha_1+...+\alpha_4$.

\begin{figure}[htb]
\centerline{\includegraphics[width=0.6\textwidth]{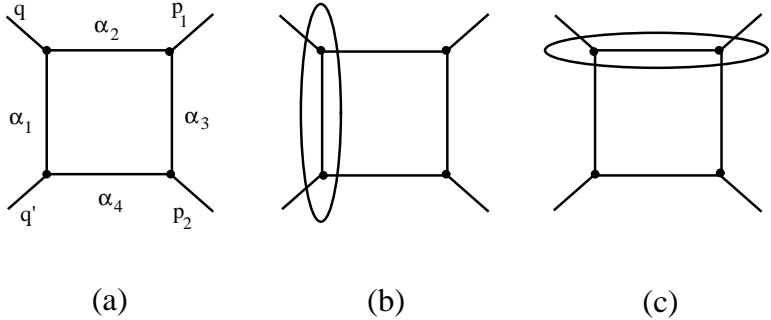}}
  \caption{(a) The box diagram in $\alpha$-space within the
           $\varphi^3_E$-theory.
           The left open lines in each graph simulate the
           ``photons'' with the appropriate kinematics (see text),
           whereas the open lines at the right denote the produced
           mock ``pions''.
           (b) ``GDA factorization'' indicated by an oval.
           (c) ``TDA factorization'' indicated by an oval.
\label{Fig01}}
\end{figure}

It is useful to introduce the dimensionless parameters
\begin{eqnarray}
\alpha_i=\frac{\tilde\alpha_i}{\Lambda^2}
\end{eqnarray}
(here, the dimensional parameter $\Lambda^2$ is an arbitrary one)
and, then, recast the $\tilde\alpha$-parameters into the form
\begin{eqnarray}
\label{rep}
  \tilde\alpha_i\biggl.\biggr|_{0}^{\infty}
=
  \frac{\beta_i}{1-\beta_i}\biggl.\biggr|_{0}^{1}.
\end{eqnarray}
As a result, we get the following expression for the considered
amplitude:
\begin{eqnarray}
\label{Amp-beta}
  &&{\cal A}(s,t,m^2)
=
  -\frac{g^4}{16\pi^2\Lambda^4}\, \tilde{\cal A} \, ,
\\
&&\tilde{\cal A}
=
  \int\limits_{0}^{1} \frac{\prod\limits_{i=1}^4 d\beta_i}{\tilde D^2}
  \text{exp} \biggl[
                    - \frac{Q^2}{\Lambda^2}
    \frac{\beta_1 \beta_2 \bar\beta_3\bar\beta_4}
         {\tilde D}
    - \frac{s}{\Lambda^2} \frac{\beta_2\beta_4 \bar\beta_1\bar\beta_3}
              {\tilde D}
    - \frac{t}{\Lambda^2} \frac{\beta_1 \beta_3 \bar\beta_2\bar\beta_4}
              {\tilde D}
    - \frac{m^2}{\Lambda^2} \frac{\tilde D}
                {\bar\beta_1\bar\beta_2\bar\beta_3\bar\beta_4}
             \biggr],
\nonumber
\end{eqnarray}
where
$
 \tilde D
=
   \beta_1\bar\beta_2\bar\beta_3\bar\beta_4
 + ... +\bar\beta_1\bar\beta_2\bar\beta_3\beta_4
$
and
$\bar\beta_i=1-\beta_i$.
The particular feature of the considered process is that the
transferred momentum $q^2=Q^2$ is large compared to the mass scale
$m^2$, the latter simulating here the typical scale of soft
interactions.
With respect to the other two kinematical variables $s$ and $t$, one
can identify four distinct regimes:
\begin{itemize}
   \item[(a)] $s\ll Q^2$ while $t$ is of order $Q^2$;
   \item[(b)] $t\ll Q^2$ while $s$ is of order $Q^2$;
   \item[(c)] $s,~t\ll Q^2$;
   \item[(d)] $s$, $t$ are both of order $Q^2$ (this regime
   corresponds to large-angle scattering \cite{Bro-Lep} and will not be
   addressed here).
\end{itemize}
Notice that within the regimes (a) and (b), amplitude (\ref{Amp1})
or (\ref{Amp-beta})} can be factorized.
Recall that the factorization of (\ref{Amp1}) and (\ref{Amp-beta}) at
the leading-twist level in the $\alpha$- or $\beta$-representation is
equivalent to the calculation of the leading $Q^2$-asymptotics of the
Laplace-type integral
\begin{eqnarray}
\label{Laplace}
   F(\lambda)=\int\limits_{0}^{\infty} d\alpha \, g(\alpha) \,
   \text{exp} \left[ - \lambda f(\alpha) \right]\approx
   \frac{g(0)}{\lambda \, f^{\prime}(0)}
   \nonumber
\end{eqnarray}
with a large and positive parameter $\lambda$ and the function
$f(\alpha)$ having a minimum at the point $\alpha = 0$.
\begin{figure}[t]
\centerline{\includegraphics[width=0.6\textwidth]{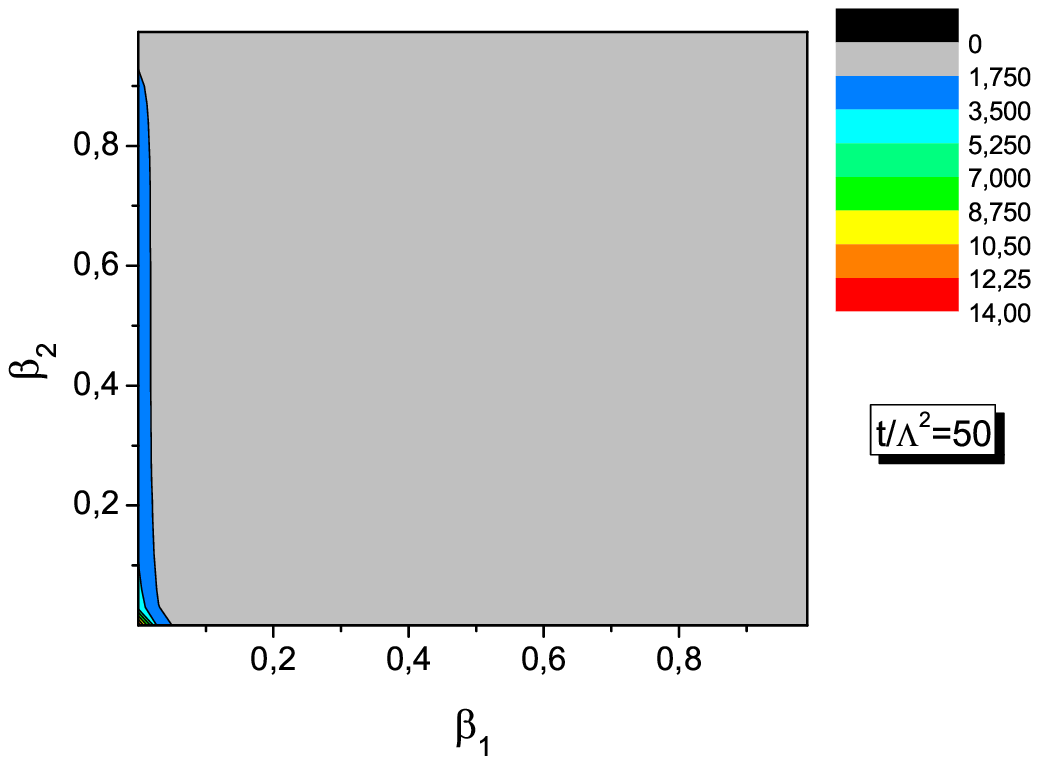}
\hspace{-1.5cm}\includegraphics[width=0.6\textwidth]{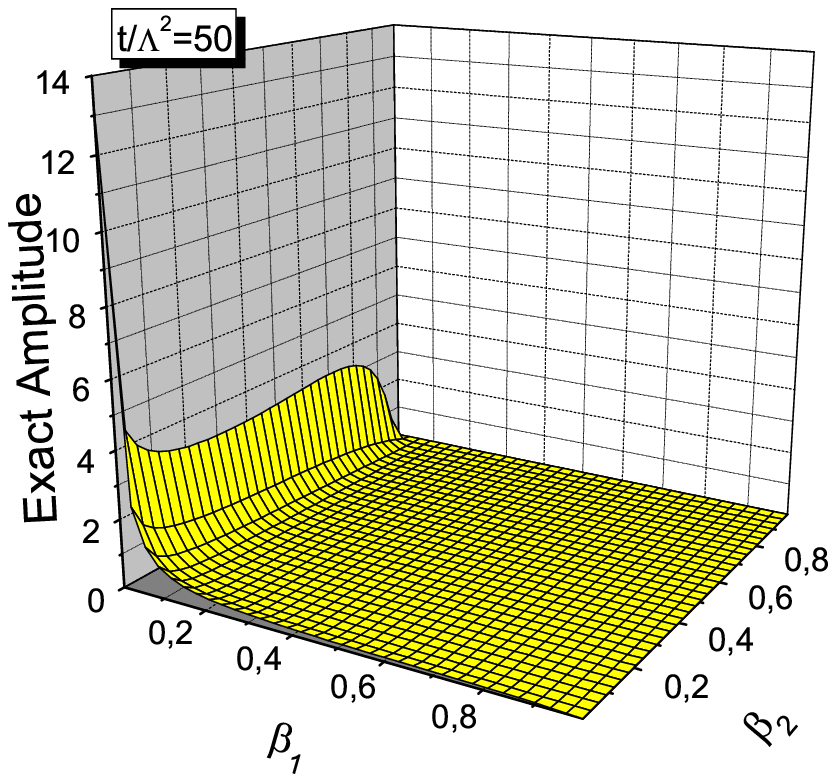}}
  \caption{The integrand of the exact amplitude (\ref{Amp-beta}) as a
  function of $\beta_1$ and $\beta_2$ in regime (a).
  Left panel: top view; right panel: 3D illustration.}
\label{Fig02}
\end{figure}

\begin{figure}[h]
\centerline{\includegraphics[width=0.3\textwidth]{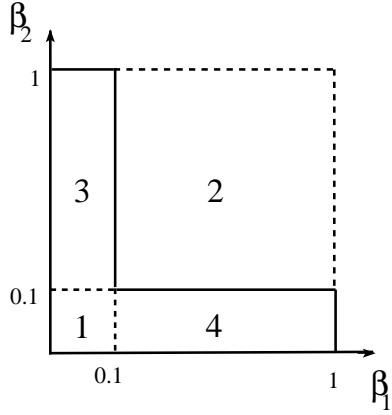}}
  \caption{Regions of integration in $\beta$ space of the integrand of
  the exact amplitude (\ref{Amp-beta}) as a function of $\beta_1$ and
  $\beta_2$.}
\label{Fig03}
\end{figure}

\textbf{Regime (a)}\\
Let us begin the discussion of this regime by considering how the
integrand in (\ref{Amp-beta}) depends on the parameters
$\beta_1$ and $\beta_2$, keeping in mind that the parameters
$\beta_3$ and $\beta_4$ will be integrated out.
To model this regime, we fix, for example, the value of $s/\Lambda^2$
to be equal to $1$ and set $t/\Lambda^2=50.0, \, Q^2/\Lambda^2=100.0$.
Then, the two-variable function representing the integrand in
(\ref{Amp-beta}) behaves as shown in Fig. \ref{Fig01}.
From this figure, one can conclude that the main contribution to the
exact amplitude (\ref{Amp-beta}) emanates from the area
restricted by small $\beta_1$ values around
$\beta_1\sim 0.05$ and $\beta_2$ values
varying in the wide interval $[0,\, 0.9]$.
We will call this support of the integrand $\beta_2$-wing.
Indeed, we may divide the whole support into the following
domains (see Fig. \ref{Fig03})
\begin{eqnarray}
\label{reg1}
&&{\rm Region-1:}\quad \beta_1\in [0,\, 0.1], \, \beta_2\in [0,\, 0.1] ;
\nonumber\\
&&{\rm Region-2:}\quad \beta_1\in [0.1,\, 1], \, \beta_2\in [0.1,\, 1] ;
\nonumber\\
&&{\rm Region-3:}\quad \beta_1\in [0,\, 0.1], \, \beta_2\in [0.1,\, 1] ;
\nonumber\\
&&{\rm Region-4:}\quad \beta_1\in [0.1,\, 1], \, \beta_2\in [0,\, 0.1]\, .
\end{eqnarray}
Our numerical computation shows that the integrations over the strip
including the regions 1 and 3 (the $\beta_2$-wing) gives
a $85\%$ contribution to the amplitude (\ref{Amp-beta}) relative to the
contribution from the whole support.
Thus, we may approximate the exact amplitude by integrating over
the $\beta_2$-wing only. [Note that the contribution from the
integration over the region 1 provides approximately a mere
5 $\%$.]

On the other hand, in this regime, the exact amplitude can
also be estimated by the asymptotic formula which results from the
analytic integration over the region where $\beta_1\sim 0$
(corresponding to the ''GDA factorization''---recall
Fig. \ref{gen-fact-mechs}):
\begin{eqnarray}
\label{AsymGDA-beta}
\tilde{\cal A}_{\text{GDA}}^{\text{as}}(s,t,m^2)
    &=&
    \int\limits_{0}^{1}
    \frac{d\beta_2 \,d\beta_3 \,d\beta_4}{\tilde D_0^2} \
    \text{exp} \left(
                     - \frac{s}{\Lambda^2}
                     \frac{\beta_2 \beta_4 \bar\beta_3}{\tilde D_0}-
                     \frac{m^2}{\Lambda^2}
                     \frac{\tilde D_0}
                          {\bar\beta_2\bar\beta_3\bar\beta_4}
               \right)
\times
\\
&&  \left[\frac{Q^2}{\Lambda^2}
    \frac{\beta_2 \bar\beta_3\bar\beta_4}{\tilde D_0}
    + \frac{t}{\Lambda^2}
      \frac{\beta_3 \bar\beta_2\bar\beta_4}{\tilde D_0}
    + \frac{m^2}{\Lambda^2}
    \right]^{-1}\, ,
\nonumber
\end{eqnarray}
where
$
 \tilde D_0
=
 \beta_2\bar\beta_3\bar\beta_4+\bar\beta_2 \beta_3 \bar\beta_4
 +\bar\beta_2\bar\beta_3 \beta_4
$.
Indeed, the integration over $\beta_1\sim 0$ eliminates the
$Q^2$-dependence in the exponential function and provides the
main contribution to the amplitude $\tilde{\cal A}$.
Schematically, this means that the propagator parameterized by
$\alpha_1$ or $\beta_1$ can be associated with the partonic (hard)
subprocesses, while the remaining propagator constitutes the soft part
of the considered amplitude, i.e., the scalar version of the GDA
(see Fig.\ \ref{Fig01}(b)).
Note that the considered process is going through the s-channel,
which would correspond to the GDA mechanism in
QCD.\footnote{In anticipation of the QCD case later on, we already use
the notions GDA and TDA, though, strictly speaking, they are not
applicable here.}

Concluding this discussion, one should stress that when
$s/\Lambda^2=1.0,\, t/\Lambda^2=50.0$,
the ratio between the asymptotic (\ref{AsymGDA-beta}) and the exact
amplitude (\ref{Amp-beta}) is
\begin{eqnarray}
\label{r1}
  R
=
  \frac{\tilde{\cal A}_{\text{GDA}}^{\text{as}}}{\tilde{\cal A}}
=
  1.01\, .
\end{eqnarray}
This ratio shows that the asymptotic formula (\ref{AsymGDA-beta})
reproduces the exact amplitude (\ref{Amp-beta}) with a rather high
accuracy, though slightly overestimating it.
Such a situation is very natural for the considered regime of the
amplitude within the $\alpha$ parametrization and is not at odds with
other factorization procedures.

\begin{figure}
\centerline{\includegraphics[width=0.6\textwidth]{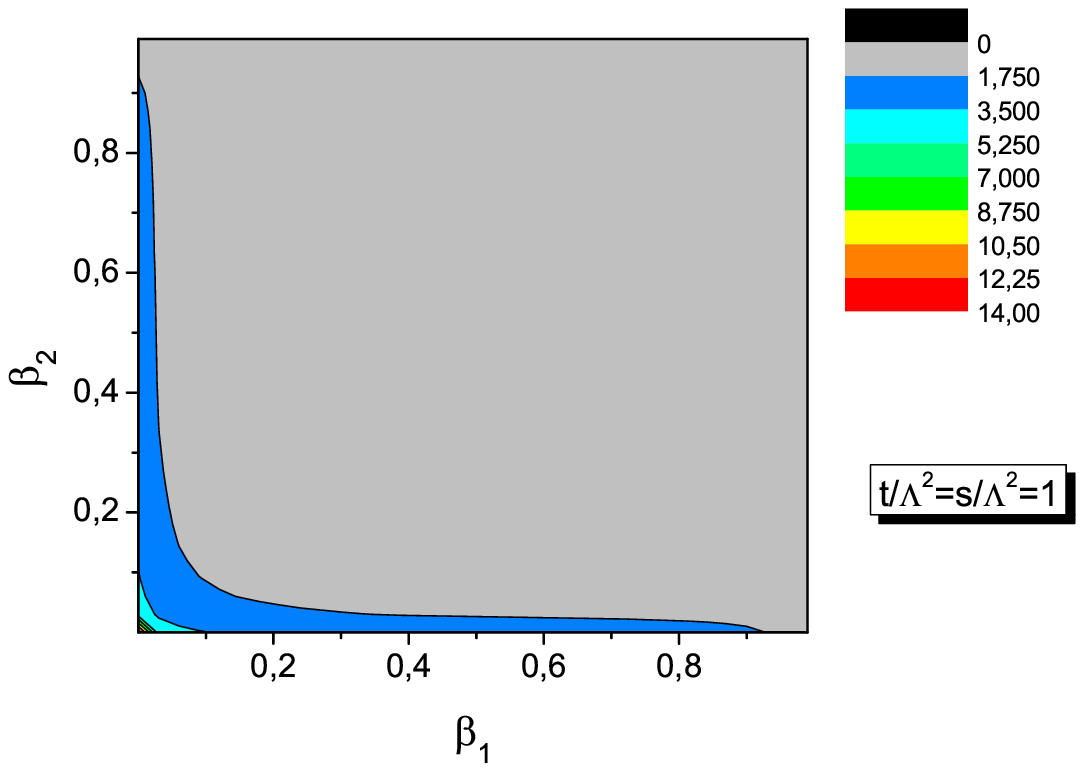}
\hspace{-1.5cm}\includegraphics[width=0.6\textwidth]{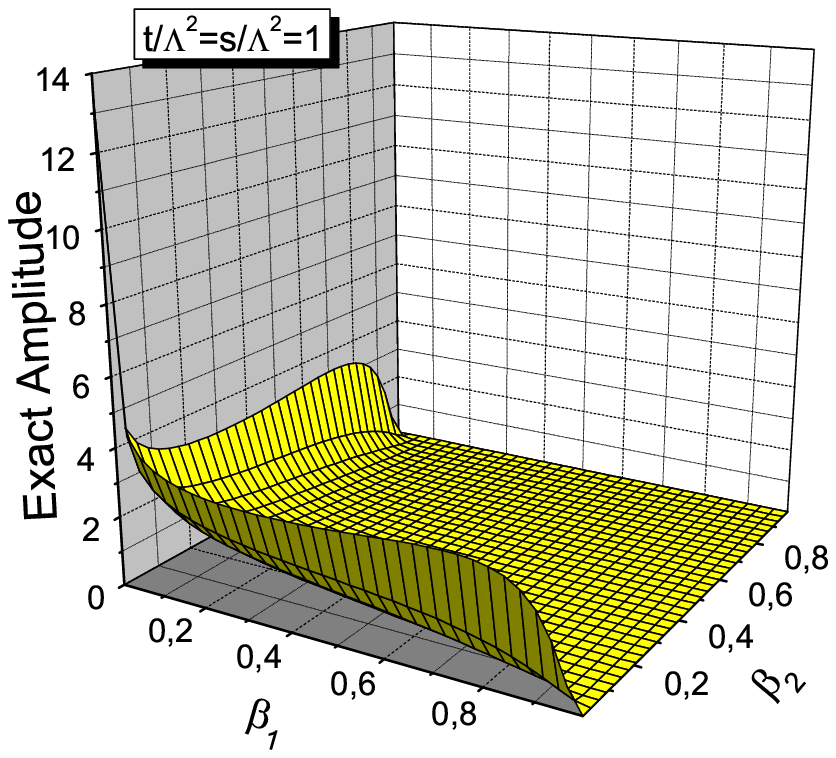}}
  \caption{The integrand of the exact amplitude (\ref{Amp-beta})
  as a function of $\beta_1$ and
  $\beta_2$ in regime (c).}
\label{Fig5}
\end{figure}

\textbf{Regime (b)}\\
Here, we have to eliminate from the exponential in Eq.\ (\ref{Amp-beta})
the variables $Q^2$ and $s$, which are large.
This can be achieved by integrating over the region $\beta_2\sim 0$,
i.e., by cutting the line corresponding to this parameter.
Performing similar manipulations as in regime (a), we find that the
scalar TDA amplitude (associated with the factorization in the
t-channel (see Fig.\ \ref{Fig01}(c))) can be related to the scalar GDA
via
\begin{eqnarray}
\label{TDAas}
  \tilde{\cal A}_{\text{TDA}}^{\text{as}}(s, t, m^2)
=
  \tilde{\cal A}_{\text{GDA}}^{\text{as}}(t, s, m^2) \, .
\end{eqnarray}

\textbf{Regime (c)}\\
The most important regime for our investigations on duality is when it
happens that both variables $s$ and $t$ are simultaneously small
compared to $Q^2$ (but still much larger than the soft scale $m^2$):
$s,\, t \ll Q^2$.
The question which may arise in this case is what
happens with factorization.
In other words, can the asymptotic formula be useful even in this
case?

In investigating this issue, we face two different options.

{\it Option 1}\\
Start with, say, the regime (a), since there is no doubt that the
asymptotic formula (or the factorization) can be applied to both
regimes (a) and (b).
We begin by considering the integrand of (\ref{Amp-beta}) in
terms of the two variables $\beta_1$ and $\beta_2$.
Then, we tune the value of the variable $t/\Lambda^2$ down to
small enough values in order to understand the behavior of
the integrand in this regime.
We originally designed the integrand of (\ref{Amp-beta})
for $t/\Lambda^2$ assuming values within the set
$\{25,\, 10,\, 5,\, 2,\, 1\}$,
but for our purposes it is necessary to study the special case
of
$t/\Lambda^2=s/\Lambda^2=1$
presented in Fig. \ref{Fig5}.
From this graphics we see that the support of the considered
function contains now two wings from which the function becomes
saturated.
Note that the $\beta_1$- and $\beta_2$-wings are the same in this
case and that both wings are equally essential.
The numerical analysis shows that the integration in (\ref{Amp-beta})
over the regions 1, 3, and 4 ($\beta_1$ and $\beta_2$ wings), i.e.,
\begin{eqnarray}
\label{Regc-exc}
\tilde{\cal A}
&\approx&
  \biggl\{
            \int\limits_{\text{Reg.-1}} +  \int\limits_{\text{Reg.-3}}
          + \int\limits_{\text{Reg.-4}}
  \biggr\}
  d\beta_1\,d\beta_2\,
  \int\limits_{0}^{1}\frac{d\beta_3\,d\beta_4}{\tilde D^2}
\\
&& \text{exp} \biggl[
                     - \frac{Q^2}{\Lambda^2}
                       \frac{\beta_1 \beta_2 \bar\beta_3\bar\beta_4}
                            {\tilde D}
                     - \frac{s}{\Lambda^2}
                       \frac{\beta_2\beta_4 \bar\beta_1\bar\beta_3}
                            {\tilde D}
                     - \frac{t}{\Lambda^2}
                       \frac{\beta_1 \beta_3 \bar\beta_2\bar\beta_4}
                            {\tilde D}
                     - \frac{m^2}{\Lambda^2}
                       \frac{\tilde D}
                            {\bar\beta_1\bar\beta_2\bar\beta_3\bar\beta_4}
              \biggr]
\nonumber
\end{eqnarray}
provides approximately a $80 \%$ contribution to the exact amplitude in
this regime.
Therefore, the saturation of the exact amplitude (\ref{Amp-beta}) in
regime (c) results from contributions originating from both wings
$\beta_1$ and $\beta_2$.
Hence, restricting attention to only one of them is not permissible.
In this sense, we deal with a particular case of ``additivity'' of
the two wings.

\begin{figure}[t]
 \centerline{\includegraphics[width=0.6\textwidth,angle=0]{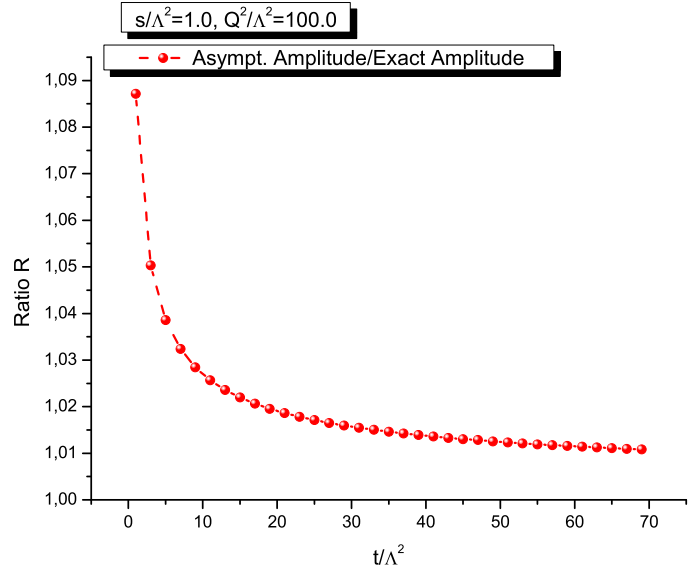}}
  \caption{The ratio $R$, see (\ref{r1}), as a function of
  $t/\Lambda^2$.}
  \label{fig-rat-1}
\end{figure}

{\it Option 2}\\
However, there exists an alternative way to determine the exact
amplitude, based on the analytic integration of (\ref{Amp-beta})
in $\beta$ space.
Namely, we may continue the asymptotic formula
(\ref{AsymGDA-beta}), obtained within regime (a),
to regime (c), where $t/\Lambda^2$ and $s/\Lambda^2$ are the
same (or approximately the same).
Then, we will observe that the asymptotic formula
(\ref{AsymGDA-beta}) still describes the amplitude
\begin{eqnarray}
\label{app-reg-c}
  \tilde{\cal A} \quad
\approx \quad
  \tilde{\cal A}_{\text{GDA}}^{\text{as}}(s,t,m^2)\biggl.
  \biggr|_{t\to s}\,
\end{eqnarray}
with an acceptable accuracy, still slightly overestimating
the exact amplitude (\ref{Amp-beta}).
This becomes evident by analyzing the ratio (\ref{r1}) in terms
of $t/\Lambda^2$, as depicted in Fig. \ref{fig-rat-1}.
One sees that the large $t$-tail of ratio (\ref{r1}) is actually
well-described by the findings in regime (a).
At the same time, when we go to the regime where $t$ is small or of the
same order as $s$, the asymptotic formula (\ref{AsymGDA-beta})---which
was continued from regime (a)---describes even in this case the exact
amplitude with a tolerable $10\%$ deviation.
Indeed, in the small $t$ region, the ratio (\ref{r1})
increases from approximately $1.01$ to about $1.09$ at
$t/\Lambda^2\approx 1$.
Hence, we may conclude that, despite the fact that the exact amplitude
(\ref{Amp-beta}) in regime (c) is saturated from contributions stemming
from both wings $\beta_1$ and $\beta_2$, the asymptotic formula
(\ref{AsymGDA-beta}), derived in regime (a) and continued to regime (c),
is still suitable for an acceptable approximation of the exact
amplitude.

Due to the symmetry of regimes (a) and (b) under the exchange of the
variables $s\leftrightarrow t$, a similar statement is valid as
regards the asymptotic formula derived in regime
(b)---cf.\ (\ref{TDAas}).
Therefore, in regime (c), the asymptotic expression
(\ref{AsymGDA-beta}) still remains a good approximation of the
exact amplitude, so that one does not have to reestimate it using as
an alternative Eq.\ (\ref{TDAas}).
Insisting to do so, one would face a ``double-counting'' problem.
Thus, in this sense, we deal with duality between (\ref{AsymGDA-beta})
and (\ref{TDAas}).
Note that this duality is not a result of the symmetry between
regimes (a) and (b).
This symmetry merely simplifies the analysis of regime (b).

Since the derivation of the asymptotic formula resembles the
factorization procedure for the given amplitude (in which the
asymptotic formula plays the role of the partonic part), the GDA and
TDA factorizations within the $\varphi_E^3$-model in regime (c) are
equivalent to each other without a kinematical or dynamical
prevalence of one over the other.
This can be interpreted as a sort of duality between the
GDA and TDA factorizations.

In concluding this discussion, let us stress that there is a
difference between the two possible ways of factorization,
which are usually considered to be equivalent, because, there is
either dominance of some parts of the full integration region,
see (\ref{Regc-exc}), or the application of the asymptotic (partonic)
formula, see (\ref{app-reg-c}) can (approximately) describe the exact
process amplitude.
In the case of regime (c), the first way of factorization implies
additivity of the two wings.
Alternatively, the continuation of the asymptotic (partonic)
expressions gives rise to duality between them.

\section{TDA-Factorization vs. GDA-Factorization for
         $\gamma\gamma^*\to\pi\pi$}
\label{sec:fact2}
\begin{figure}[b]
 \centerline{\includegraphics[width=0.6\textwidth,angle=0]{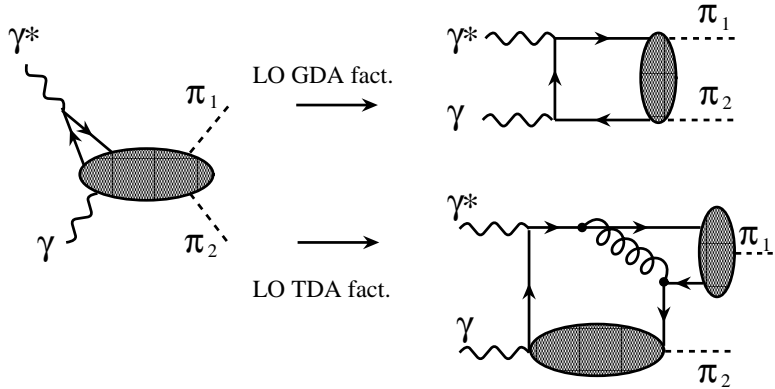}}
\vspace{0.0cm}
  \caption{Two ways of factorization: (Left panel) via the GDA mechanism
  and via the TDA mechanism (right panel).
  Only a single typical diagram for each case is shown for the
  purpose of illustration.}
  \label{GDAvsTDA}
\end{figure}

In the preceding section, we demonstrated how the duality between the
GDA and the TDA-factorization works within a toy model.
We now study the duality phenomenon in the case of real QCD.
To this end, we consider the exclusive $\pi^+\pi^-$ production in a
$\gamma_T\gamma^*_L$ collision, where the virtual photon with a large
virtuality $Q^2$ is longitudinally polarized, whereas the other one is
quasi real and transversely polarized.
As already mentioned, the key feature of this process is that it exhibits
two different kinds of factorization, based, respectively, on the GDA or
the TDA mechanism (see Fig. \ref{GDAvsTDA}), with a
potential overlap of both mechanisms in the kinematical regime
where $t$ and $s$ are small.
Notice that this situation differs from the cases studied in
\cite{PSSW}, where the additivity of the GDA and the TDA factorizations
was explored and found that each factorization mechanism can be
selected on account of the helicity of the particles.

In contrast, in our case, the GDA and the TDA regimes correspond to
the \textit{same} helicity amplitudes, resembling the situation faced
in the scalar case considered in the preceding section where there is
only one helicity amplitude.
To continue, let us recall the parametrization of nonperturbative
matrix elements entering our analysis.
Because the considered process involves a longitudinally and a
transversally polarized photon, we have to deal with twist-3
parametrizing functions.
To this end, we adhere to \cite{AT-WW}, where the genuine twist-$3$
contributions in the $\gamma\gamma^*\to\pi\pi$ process were studied
in detail (see, also, \cite{Kiv}).
As regards the twist-2 contribution, related to the meson DA,
we use the standard parametrization of the $\pi^+$-to-vacuum matrix
element which involves a bilocal axial-vector quark operator
\cite{Efremov-Radyushkin}.

On the other hand, the $\gamma\to\pi^-$ matrix elements, entering the
TDA-factorized amplitude, can be parameterized in the form
(cf. \cite{Pire-Szym})
\begin{eqnarray}
&& \langle \pi^-(p_2)| \bar \psi(-z/2)\gamma_\alpha [-z/2;z/2]\psi(z/2)
   |\gamma(q^\prime, \varepsilon^\prime)
   \rangle \stackrel{{\cal F}}{=}
   \frac{ie}{f_\pi}\epsilon_{\alpha\varepsilon^\prime_T P \Delta_T}
   V_1(x,\xi,t)\,,
   \nonumber \\
&& \langle \pi^-(p_2)| \bar \psi(-z/2)\gamma_\alpha \gamma_5
   [-z/2;z/2]\psi(z/2)
   |\gamma(q^\prime, \varepsilon^\prime)
   \rangle \stackrel{{\cal F}}{=}
   \frac{e}{f_\pi}\varepsilon^\prime_T
   \cdot\Delta_T P_\alpha A_1(x,\xi,t)\, ,
\label{eq:gpimeA}
\end{eqnarray}
where  the symbol $\stackrel{{\cal F}}{=}$
stands for the Fourier transformation (with the appropriate measure)
and
\begin{equation}
 [-z/2;z/2]
=
 {\cal P}\exp \left[
                    ig\int_{z/2}^{-z/2} d x_{\mu}
                    A_{a}^{\mu}(x)t_{a}
              \right]
\label{eq:link}
\end{equation}
denotes a path-ordered gauge link to ensure gauge invariance.
In the following, we choose to work in the axial gauge which entails
that the gauge link is equal to unity.
In (\ref{eq:gpimeA}), the nonperturbative amplitudes $V_1$ and $A_1$
denote, respectively, vector and axial-vector TDAs, whereas the
relative momentum is defined as
$P=(p_2+q^\prime)/2$, and $\Delta=p_2-q^\prime$.
Notice that TDAs in the ERBL region, i.e., $-\xi\leq x \leq\xi$,
include the so-called D-term, where the pole contribution, related to
the spinless resonance (this being the pion in our case), is included
\cite{Pol-NPB,Pol-Wei}.
Finally, to normalize the axial-vector TDA, $A_1$, we adopt the
philosophy of
\cite{Pire-Szym,Prash,Tib05,BA07,CN07,AnDor1,AnDor2} and express
$A_1$ in terms of the axial-vector form factor measured in the
weak decay $\pi\to l\nu_l \gamma$, i.e.,
\begin{eqnarray}
\label{normA}
   \int\limits_{-1}^1 dx \, A_1(x,\,\xi,\, t)
   = 2\, f_\pi \, F_{A}(t)/m_{\pi}\,  ,
\end{eqnarray}
where $f_\pi=0.131\, \text{GeV}$, $m_\pi=0.140\, \text{GeV}$, and
$F_A(0)\approx 0.012$ \cite{PDG06,Byc08}.

\begin{figure}[bt]
 \centerline{\includegraphics[width=0.7\textwidth,angle=0]{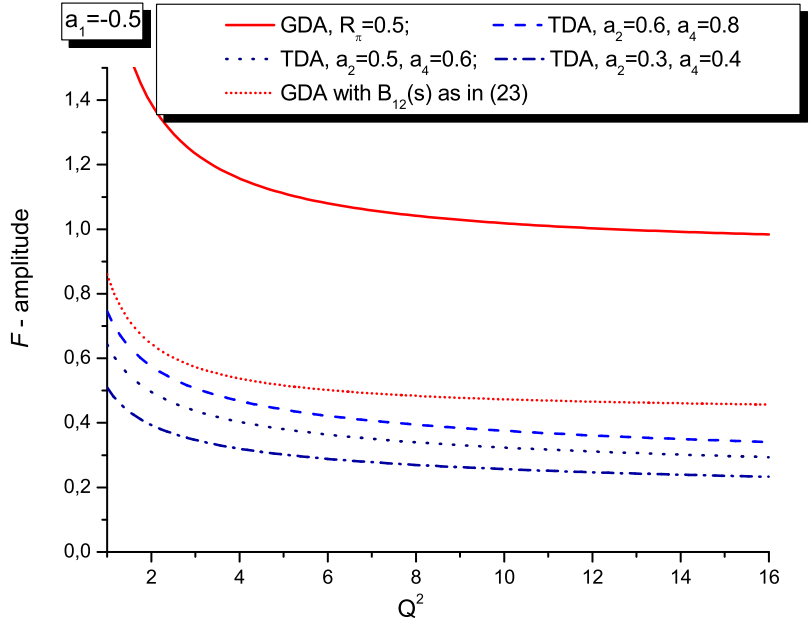}}
  \caption{Helicity amplitudes ${\cal F}^{\text{TDA}}$ [cf.\ Eq.\
  (\ref{eq:haTDA})] and ${\cal F}^{\text{GDA}}$ [cf.\ Eq.\
  (\ref{eq:amp_GDA_WW}] as functions of $Q^2$, using $a_1=-0.5$
  found in lattice simulations \protect{\cite{Lattice}.}
  The value of $s/Q^2$ varies in the interval
  $[0.06,\, 0.3]$.
}
  \label{fig-TDA-GDA1}
\end{figure}

The next objects of interest in our considerations are the helicity
amplitudes that are obtained from the usual amplitudes after
multiplying them by the photon polarization vectors:
\begin{eqnarray}
   {\cal A}_{(0,\pm)}=\varepsilon^{(0)}_{\mu}
   T^{\mu\nu}_{\gamma\gamma^*}\varepsilon^{\,\prime\,(\pm)}_{\nu},
   \quad
   \varepsilon^{\,\prime\,(\pm)}
   =
   \left( 0,\frac{\mp 1}{\sqrt{2}},\frac{+i}{\sqrt{2}},0 \right),
   \quad
   \varepsilon^{\,(0)}=
   \left(\frac{|\vec{q}\,|}{\sqrt{q^2}},0,0,\frac{q_0}{\sqrt{q^2}}
   \right) \, .
   \label{eq:hel}
\end{eqnarray}
Discarding the $\Delta^2_T$ corrections, the only remaining
contribution stems from the axial-vector one ($A_1$) given by Eq.\
(\ref{eq:gpimeA}).
Thus, the helicity amplitude associated with the TDA mechanism reads
\begin{eqnarray}
\label{TDAhelam}
   {\cal A}^{\text{TDA}}_{(0,j)}
   = \frac{\varepsilon^{\prime\,(j)}
   \cdot\Delta^T}{|\vec{q}\,|} {\cal F}^{\text{TDA}}, \quad
   {\cal F}^{\text{TDA}}=
   [4\,\pi\,\alpha_s(Q^2)]\frac{C_F}{2\,N_c}
   \int\limits_{0}^{1} dy \, \frac{\phi_\pi(y)}{y\bar y}
   \int\limits_{-1}^{1} dx \, A_1(x,\xi,t)\,
   \biggl( \frac{e_u}{\xi-x}-\frac{e_d}{\xi+x} \biggr),
   \label{eq:haTDA}
\end{eqnarray}
where
\begin{eqnarray}
\label{xi}
\xi^{-1}=1+\frac{2 W^2}{Q^2} \, ,
\end{eqnarray}
and where we have employed the 1-loop $\alpha_s(Q^2)$
in the $\overline{\rm MS}$-scheme with $\Lambda_{\rm QCD}=0.312$~GeV
for $N_f=3$ \cite{Kataev:2001kk}.
Note that there is only a mild dependence on $\Lambda_{\rm QCD}$.
Choosing, for instance, a somewhat smaller value, say, around
$0.200$~GeV, would shift the TDA curves shown in Fig.\
\ref{fig-TDA-GDA1} slightly downwards.

On the other hand, the helicity amplitude which includes the
twist-$3$ GDA can be written as (see, for example, \cite{AT-WW,Kiv})
\begin{eqnarray}
\label{GDAhelam}
   {\cal A}_{(0,j)}^{\text{GDA}}
   =\frac{\varepsilon^{\,\prime\,(j)}\cdot\Delta^T}{|\vec{q}\,|}
   {\cal F}^{\text{GDA}}, \quad
   {\cal F}^{\text{GDA}}=2 \frac{W^2+Q^2}{Q^2}
   (e^2_u+e^2_d)
   \int\limits_{0}^{1} dy \, \partial_{\zeta} \Phi_1(y,\zeta,W^2)
   \biggl( \frac{\ln{\bar y}}{y} - \frac{\ln{y}}{\bar y}\biggr)\, ,
   \label{eq:amp_GDA_WW}
\end{eqnarray}
where
\begin{eqnarray}
\label{zeta}
  2\zeta-1=\beta\cos{\theta_{cm}^\pi} ,
\quad
  \cos{\theta_{cm}^\pi}=\frac{2t}{W^2+Q^2} - 1
\end{eqnarray}
and the partial derivative is defined by
$\partial_\zeta = \partial/\partial(2\zeta-1)$.
Here the amplitude $\Phi_1(y,\zeta,W^2)$ encodes the matrix element
of the twist-2 operator.
In deriving (\ref{GDAhelam}), we have used the Wandzura-Wilczek
approximation for the twist-$3$ contribution and took into account
that in our system $(p_2-p_1)^T=2(p_2-q^\prime)^T$.
Besides, we would like to stress that the amplitudes (\ref{TDAhelam})
and (\ref{GDAhelam}) are both of order $1/Q^2$, which offers room for
duality.
As pointed out above, duality between expressions (\ref{eq:haTDA})
and (\ref{eq:amp_GDA_WW}) may occur in that regime where both
Mandelstam variables $s$ and $t$ are much smaller in comparison to the
large photon virtuality $Q^2$.
In terms of the skewness parameters $\xi$ and $\zeta$, this would
imply that $\xi\sim 1$ and $\zeta\sim 0$.

To estimate the relative weight of the amplitudes with TDA or GDA
contributions, we have to model these non-perturbative objects.

Let us start with the TDAs.
As a first step, we assume a factorizing ansatz for the $t$-dependence
of the TDAs and write
\begin{eqnarray}
\label{t-dep}
   A_1(x,\xi,t)=  2\, \frac{f_\pi}{m_\pi} \, F_{A}(t) A_1(x,\xi) \, ,
\end{eqnarray}
where the $t$-independent function $A_1(x,\xi)$ is normalized to
unity:
\begin{eqnarray}
\label{normA-unity}
   \int\limits_{-1}^{1} dx A_1(x,\xi)=1 \, .
\end{eqnarray}
To satisfy this condition, we introduce a TDA defined by
\begin{eqnarray}
    A_1(x, 1)
    =\frac{A_1^{\text{non-norm}}(x,1)}{{\cal N}}, \quad {\cal N}
    =\int\limits_{-1}^1 dx A_1^{\text{non-norm}}(x,1) \, .
\end{eqnarray}

We now focus on the discussion of the $t$-independent TDAs in
(\ref{t-dep}).
Since we are mainly interested in TDAs in the
Efremov-Radyushkin-Brodsky-Lepage (ERBL)
\cite{Efremov-Radyushkin,Bro-Lep} region $\xi=1$, it is instructive
to choose the following parametrization
\begin{eqnarray}
\label{TDAansatz}
    A_1^{\text{non-norm}}(x, 1)
    =
      (1-x^2)\biggl( 1+ a_1 C^{(3/2)}_1(x)
    + a_2 C^{(3/2)}_2(x)+ a_4 C^{(3/2)}_4(x)\biggr) ,
\end{eqnarray}
where $a_1, \,a_2, \, a_4$ are free adjustable parameters, encoding
nonperturbative input, and the standard notations for Gegenbauer
polynomials are used.
One appreciates that the TDA in the form of Eq.\ (\ref{TDAansatz})
amounts to summing a $D$-term, \textbf{i.e.,} the term with the
coefficient $a_1$, and meson-DA-like contributions.
Indeed, would we eliminate the term with $a_1$, we would obtain the
standard parametrization for a meson DA.
On the other hand, keeping only the term with $a_1$, would reproduce
the parametrization for the $D$-term \cite{GPV}.
Therefore, for our analysis, we suppose that
$a_1\equiv d_0$ \cite{GPV}, which is equal to $-0.5$ in lattice
simulations \cite{Lattice}.\footnote{ The case of $a_1=-4/ N_f$,
which holds within the chiral quark-soliton model \cite{CQM}, is
also included in our analysis.
We found that there is no much difference in comparison with the
lattice model.}
With respect to the parameters $a_2$ and $a_4$, we allow them to vary
in quite broad intervals, notably,
$a_2\in [0.3, \, 0.6]$ and $a_4\in [0.4, \, 0.8]$,
that would cover vector-meson DAs with very different
profiles at a normalization scale $\mu^2\sim 1\, {\text{GeV}^2}$
(see, for example, \cite{BM-rhomes}).
Note that the profiles obtained with ansatz (\ref{TDAansatz}) bear
remarkable resemblance to the TDAs calculated within a
non-local chiral
quark-soliton model in the recent work of Ref.\ \cite{Prash}.

As regards the function $\Phi_1(z,\zeta)$, which denotes the
corresponding GDA in (\ref{eq:amp_GDA_WW}), we take recourse to the
following model \cite{Diehl:2003}
\begin{eqnarray}
\label{Phi1}
    \Phi_1(z, \zeta, W^2)=9 N_f\, z \bar z (2z-1)\,
    \left( \tilde B_{10}(W^2) e^{i\delta_0(W^2)}
           +  \tilde B_{12}(W^2) e^{i\delta_2(W^2)} P_2(\cos\theta_\pi)
    \right)\, ,
\end{eqnarray}
where the phase shift of the $\pi\pi$ scattering is defined by
$\delta_0(W_0=0.8)\approx \frac{\pi}{2}$ and
$\delta_2(W_0=0.8)\approx 0.03 \pi$ \cite{DGP,DGPT98}.
In (\ref{Phi1}), the function $\tilde B_{10}$ corresponding to $L=0$
does not contribute to (\ref{eq:amp_GDA_WW}) and, therefore, can be
discarded.
On the other hand, the function $\tilde B_{12}$ can be estimated by the
method described in \cite{APSTW-Hyb}.
Since $\tilde B_{12}$ corresponds to the two-pion state with $L=2$
and our GDAs probe only the isoscalar channel,
we derive for $\tilde B_{12}$ at $W^2$ in the
region of the $f_2$-meson [$I^G(J^{PC})=0^+(2^{++})$] mass
the following expression:
\begin{eqnarray}
\label{B12}
     \tilde B_{12}(W^2)=\frac{10}{9}\frac{g_{f_2\pi\pi}\,
     f_{f_2}\, M^3_{f_2}\,\Gamma_{f_2}}
     {(M_{f_2}^2-W^2)^2+\Gamma^2_{f_2}M^2_{f_2}} \, ,
\end{eqnarray}
where $f_{f_2}=0.056\, \text{GeV}, \, M_{f_2}=1.275\, \text{GeV}, \,
\Gamma_{f_2}=0.185\, \text{GeV}$,
and the decay constant $g_{f_2\pi\pi}$ is defined by
\begin{eqnarray}
   g_{f_2\pi\pi}=\sqrt{\frac{24\pi}{M^3_{f_2}} \,
   \Gamma(f_2\to \pi\pi)}, \quad
   \Gamma(f_2\to \pi\pi)=0.85 \Gamma_{f_2} \, .
\end{eqnarray}
In addition, we also model the function $\tilde B_{12}$ with
the help of the most simple ansatz (see, \cite{DGP})
\begin{eqnarray}
\label{Bsimp}
     \tilde B_{12}(0)=\beta^2 \frac{10}{9 N_f} R_\pi \, ,
\end{eqnarray}
where $R_\pi$ denotes the fraction of momentum carried by the quarks
and antiquarks in the pion.
In our numerical analysis below, we use for this
fraction $R_\pi=0.5$, which is taken from the leading-order
Gl\"{u}ck-Reya-Schienbein approximation (see, for example, \cite{DGP}).

Before continuing with the numerical analysis, let us pause for a
moment to discuss the validity and self-consistency of the
approximations made.
Start with the pion DA: For simplicity, the asymptotic pion DA in Eq.\
(\ref{eq:haTDA}) was used.
Using other pion DAs \cite{BMS05lat}, the contribution from the
TDA factorization may increase from about $6\%$ for the BMS model
\cite{BMS01} to approximately $56\%$ for the CZ model \cite{CZ84}.
Second, continue with the GDA: At the level of accuracy relevant for
our arguments, we used for the twist-2 GDA $\Phi_1$ the asymptotic
form, the justification being provided by quark-hadron duality
arguments given in \cite{Rad95}.
Third, we assumed the Wandzura-Wilczek approximation, which relies
upon the assumption of the smallness of the genuine twist-3
contributions.
This assumption is supported by measurements of the $g_2$ structure
function.
The normalization of the GDA is determined by the momentum conservation
which is rigorously valid.
By performing a crossing from GPDs (where momentum conservation
applies) to the GDA channel, we used the improved expression
(\ref{B12}) accounting for meson dynamics.
Turn to the TDA: The first ingredient is the normalization
(\ref{normA}), which is fixed by the data for the pion weak decay
(see for details \cite{Pire-Szym}).
Finally, for the form of the TDA, we used a parametrization
[cf.\ (\ref{TDAansatz})], which can accommodate different available
models \cite{Tib05,BA07,CN07,Prash}, within the range of accuracy of
the order of 10\%, the latter pertaining to our whole numerical
analysis.

We calculated both functions
${\cal F}^{\text{TDA}}$ and ${\cal F}^{\text{GDA}}$
and show the obtained results in Fig.\ \ref{fig-TDA-GDA1}
for $a_1= -0.5$ (which corresponds to the lattice result of
\cite{Lattice}) and employing for simplicity the asymptotic pion DA.
The dashed line corresponds to the function ${\cal F}^{\text{TDA}}$,
where we have adjusted the free parameters to $a_2=0.6$ and $a_4=0.8$.
The results, obtained for rather small values of these parameters, are
displayed by the broken lines in the same figures.
The dotted line denotes the function ${\cal F}^{\text{TDA}}$ with
$a_2=0.5$ and $a_4=0.6$, whereas the dash-dotted line employs
$a_2=0.3$ and $a_4=0.4$.
For comparison, we also include the results for the
${\cal F}^{\text{GDA}}$.
In that latter case, the dense-dotted line corresponds to the GDA
amplitude, where the expression for $\tilde B_{12}$ has been estimated
via Eq.\ (\ref{B12}), while the solid line represents the simplest
ansatz (\ref{Bsimp}) with $R_\pi=0.5$.
The particular values of the coefficients $a_2, a_4$, we used, are
within the fiducial intervals specified above in conjunction with Eq.\
(\ref{TDAansatz}).
If, for example, we would have used instead values of these parameters
outside their fiducial intervals, then the outcome would change by
orders of magnitude and duality would be badly violated.

From this figure one may infer that (i) in the case when the parameter
$\tilde B_{12}$ (which parametrizes the GDA contribution) is estimated
with the aid of the Breit-Wigner formula, see (\ref{B12}), and,
while $s,\,t \ll Q^2$, there is duality between the GDA and the
TDA factorization mechanisms.
Hence, the model for $\tilde B_{12}$ which takes into account the
corresponding resonances can be selected by duality.
(ii) In the case of the simplest model for $\tilde B_{12}$, evaluated
for $R_\pi=0.5$ (see, (\ref{Bsimp})), the duality between the GDA and
the TDA factorization mechanisms reaches maximally the level of
$35 \, \%$, provided the axial-vector TDAs are normalized in terms of
the axial form factor $F_A$.
On that basis, one may conclude that the simplest model for
$\tilde B_{12}$ is rather inconsistent with duality
and cannot be considered as a realistic option.

\section{Summary and Conclusions}
\label{sec:concl}

The process amplitude for the fusion of a highly virtual, but small
$s$ photon (longitudinally polarized), with a quasi real one (with
a transversal polarization) can be factorized in two different
ways.
The study of this process within a toy $\varphi_E^3$ model has shown
the existence of duality between the factorization along the
$s$-channel and the $t$-channel---which serve to simulate,
respectively, the GDA and the TDA factorization mechanisms in QCD.
This observation persists also in the case of such reactions in
real QCD.
Indeed, the amplitude for the process $\gamma\gamma^*\to\pi\pi$ can be
factorized in two distinct ways, having recourse to the large
virtuality of one of the photons involved, as compared to the
scales of the soft interactions.
As mentioned before, depending on the variables $s$ and $t$, two
different mechanisms for this reaction were identified.
The first factorization mechanism, which employs GDAs, takes place when
$s\ll Q^2$, while $t$ is of the order of $Q^2$.
On the other side, when it happens that $t\ll Q^2$ and $s\sim Q^2$,
the TDA mechanism of factorization becomes dominant in such a
$\gamma\gamma^*$ collision.
We have shown that when both variables $s$ and $t$ are simultaneously
much smaller than $Q^2$ in the collision of a real and transversely
polarized photon with a highly virtual longitudinally polarized
photon in the pion-pair production, then both types of factorization
are possible giving rise to duality in this region.

Our observations in QCD can be summarized
and outlined as follows.

We have shown that when it happens that both Mandelstam variables
$s$ and $t$ are much less than the large momentum scale $Q^2$,
with the variables $s/Q^2$ and $t/Q^2$ varying
in the interval $(0.001, \, 0.7)$, then the TDA and the GDA
factorization mechanisms are equivalent to each other and operate in
parallel.
This marks a crucial difference between our approach and previous ones
\cite{PSSW} which dealt with the additivity of the TDA and the GDA
factorization mechanisms.
Note that the presented results were obtained using rigorous
constraints and available nonperturbative models for the DAs, GDAs,
and the TDAs.
We found that duality appears only for values of the model
parameters within their fiducial intervals, associated with each
particular model.
Therefore, we demonstrated that duality may serve as a tool for
selecting suitable models for the non-perturbative ingredients
of QCD factorization of various exclusive amplitudes.

We also observed that twist-3 GDAs appear to be dual to the
convolutions of leading-twist TDAs and DAs, multiplied by a QCD
effective coupling---the latter stemming from the hard-gluon exchange.
This reflects at the level of the particular process, we are
considering, the manifestation of duality for the GPDs and GDAs
themselves \cite{Pol-NPB,Polyakov:2002wz,Moiseeva:2008qd}.
One may ask to which extent the observed duality between different QCD
factorization mechanisms is an exceptional finding or rather represents
a generic feature of QCD dynamics.
In this connection, one may recall the different ways of describing
single-spin asymmetries (SSA) using either twist-3 collinear
factorization \cite{Efremov:1984ip,Qiu:1991pp} or by employing the
transverse-momentum dependent (TMD) Sivers distribution function
\cite{Sivers:1989cc}.
Because of the T-odd nature of the SSA and the Sivers function, the
latter may emerge only due to the interactions between the hard and the
soft parts \cite{Brodsky:2002cx} of the process and represents an
effective distribution \cite{Boer:1997bw}.
Moreover, its transverse moment is related \cite{Boer:2003cm} to a
particular twist-3 matrix element.
This sort of relation may be explored \cite{Ji:2006ub} as a matching
procedure (or, in our language, as duality) between TMD factorization
at low $P_T$, on one hand, and collinear twist-3 factorization at
large $P_T$, on the other.
Recently \cite{Bacchetta:2008xw}, such a matching procedure was carried
out in detail for various observables and it was found that duality
strongly depends on the type of the considered observable.
In the case of the Sivers function, there is a possibility of enlarging
the duality intervals by either using the twist-3 approach at low $P_T$
\cite{Teryaev:2005bp} or by applying the Sivers function at large $P_T$
\cite{Ratcliffe:2007ye}.
The latter procedure finds justification in the framework of the
twist-3 factorization, provided only the first non-vanishing transverse
moment of the Sivers function is retained and that color factors,
responsible for the effective nature of the Sivers function, are
properly taken into account.

Duality also emerges when another type of QCD factorization in
semi-inclusive processes is considered, which makes use of so-called
fracture functions.
The momentum conservation guarantees that the description in terms of
either fragmentation or fracture functions is complete and one should
have duality between these two mechanisms in the overlap region (see
\cite{Teryaev:2004df} and references therein).
Bottom line:
We have shown that duality leads to new insights into a number of
phenomena in QCD, in particular to those related to exclusive
processes and spin-dependent quantities.

\subsection*{Acknowledgments}

We would like to thank A.B.~Arbuzov, A.~P.~Bakulev, A.E.~Dorokhov, A.~V.~Efremov, N.~Kivel,
B.~Pire, M.~V.~Polyakov, M.~Prasza{\l}owicz, L.~ Szymanowski, and
S.~Wallon for useful discussions and remarks.
This investigation was partially supported by the Heisenberg-Landau
Programme (Grant 2008), the Deutsche Forschungsgemeinschaft under
contract 436RUS113/881/0, the Alexander
von Humboldt-Stiftung, the EU-A7 Project \emph{Transversity},
the RFBR (Grants 06-02-16215,08-02-00896 and 07-02-91557),
the Russian Federation Ministry of Education and Science
(Grant MIREA 2.2.2.2.6546), the RF Scientific Schools grant 195.2008.9,
and INFN.



\end{document}